\documentclass[12pt]{article}
\usepackage{epsfig}

\def\be{\begin{equation}}

\def\ee{\end{equation}}

\def\bea{\begin{eqnarray}}

\def\eea{\end{eqnarray}}

\newcommand{\ba}{\begin{array}}
\newcommand{\ea}{\end{array}}

\newcommand{\doublespace}{
    \renewcommand{\baselinestretch}{1.6}\large\normalsize}

  \advance\oddsidemargin  by -1in
  \advance\evensidemargin by -1in
  \advance\topmargin      by -0.5in
\def\lsim{\mathrel{\rlap{\lower4pt\hbox{\hskip1pt$\sim$}}
    \raise1pt\hbox{$<$}}}         
\def\gsim{\mathrel{\rlap{\lower4pt\hbox{\hskip1pt$\sim$}}
    \raise1pt\hbox{$>$}}}         

\def\Pom{{\bf I\!P}}
\def\beq{\begin{equation}}
\def\endeq{\end{equation}}
\def\arr{\begin{eqnarray}}
\def\endarr{\end{eqnarray}}
\makeindex

\doublespace

\begin{document}


\phantom{.}{\bf \Large \hspace{8.0cm} KFA-IKP(Th)-1996-06 \\
\phantom{.}\hspace{11.9cm}July 1996\vspace{0.4cm}\\ }

\begin{center}
{\bf\sl \huge
Diffractive DIS: back to triple-Regge phenomenology?}
\vspace{0.4cm}\\
{\bf
N.N.Nikolaev$^{a),b),c)}$, W.Sch\"afer$^{b)}$, B.G.Zakharov $^{c)}$
}
\vspace{.5cm}\\
{\sl
$^{a)}$ITKP der Universit\"at Bonn, Nu{\ss}allee 14-16, D-53115 Bonn\\
$^{b)}$IKP, KFA J\"ulich, D-52425 J\"ulich, Germany\\
$^{c)}$L.D.Landau Institute, Kosygina 2, 1117 334 Moscow, Russia}
\vspace{1.0cm}\\
{\Large
Abstract}\\
\end{center}
We discuss the factorization breaking effects caused by
the contribution to large rapidity gap events from
DIS on secondary reggeons. Based on the triple-Regge
phenomenology of hadronic diffraction dissociation, we
present estimates for the flux and structure function of
the $f$ reggeon. The kinematical $x_{\Pom}$--$\beta$
correlations is shown to modify substantially the observed
$x_{\Pom}$ dependence of the diffractive structure function.
The secondary reggeon and $x_{\Pom}$--$\beta$ correlation
effects explain the recent H1 finding of the factorization
breaking and resolve the apparent contradiction between
the preliminary H1 results and predictions from the color dipole
gBFKL approach. We suggest further tests of predictions
for diffractive DIS from the gBFKL approach.
\bigskip\\

\begin{center}
E-mail: kph154@ikp301.ikp.zam.kfa-juelich.de
\end{center}

\pagebreak

\section{Introduction: rapidity gaps from non-pomeron exchanges?}

There is great interest in diffractive DIS $\gamma^{*}+p
\rightarrow X +p'$ as a probe of the QCD pomeron.
A convenient quantity is a diffractive structure
function operationally defined as \cite{Regge,Ingelman,DLPom,NZ92}
\arr
\left.(M^{2}+Q^{2})
{ d\sigma^{D} (\gamma^{*}\rightarrow X)
\over dt\,d M^{2} }\right|_{t=0} =
{  \sigma_{tot}(pp) \over 16\pi}
{4\pi^{2} \alpha_{em}
\over Q^{2}}
 F^{D}(x_{\Pom},\beta,Q^{2})\, .
\label{eq:1}
\endarr
Here $Q^{2}$ is the virtuality of the photon, $W$ and $M$
are c.m.s. energy in the photon-proton and photon-pomeron
collision, $\beta =Q^{2}/(Q^{2}+M^{2})$ has the meaning of
the Bjorken variable for the lepton-pomeron DIS,
$x_{\Pom}=(Q^{2}+M^{2})/(Q^{2}+W^{2})=x/\beta$ is
interpreted as the fraction of the momentum of the proton
carried away by the pomeron, $t$ is the $p$-$p'$ momentum
transfer squared. Of special interest is the $x_{\Pom}$-dependence
of $F^{D}$, which measures the spin $j$ (intercept) of the object
exchanged in the $t$ channel: $F_{D} \propto x_{\Pom}^{2(1-j)}$.
Color dipole gBFKL dynamics, one of the successful approaches to
LRG (large rapidity gap) physics, predicts  
\cite{NZ92,NZ94,GNZ95,GNZcharm,GNZlong} that
at the moderately small values of $x_{\Pom}$ presently accessible at
HERA the exponent $n=2j-1$ must depend on flavor, $\beta$
and, for longitudinal photons, on $Q^{2}$  in
defiance of the often assumed Ingelman-Schlein-Regge
factorization \cite{Regge,Ingelman,DLPom,Capella,Kwiecinski,Stirling}.
The rise of the exponent $n$ at small $x_{\Pom}$ is of
special interest as it derives from the
intrusion of hard scattering effects into soft interaction
amplitudes which is a very specific signature of the gBFKL
dynamics \cite{BFKL,gBFKL,NZHera}.

One of the signals of the Regge factorization breaking in the gBFKL
approach is a {\sl rise} of $n(x_{\Pom},\beta)$ towards small
$\beta $ \cite{GNZ95}. Recently the H1 collaboration reported
the first evidence for the factorization breaking. However, H1
finds a {\sl decrease} of  $n(x_{\Pom},\beta)$ at small $\beta$
\cite{H1Roma}. Is that compatible with the gBFKL approach?

Although only the pomeron exchange survives at $x_{\Pom} \rightarrow 0$,
the range of the presently accessible $x_{\Pom}$
is limited. Furthermore, by virtue of the kinematical relationship
$x_{\Pom}=x/\beta$, the small-$\beta$ data correspond
to the larger values of $x_{\Pom}$. Although the confirmation
of the H1 effect by the ZEUS collaboration is pending \cite{ZEUSRoma},
one must seriously examine the possibility \cite{NZRoma,LandshoffRoma}
that the H1 effect is due to an admixture of the non-pomeron
exchanges. Indeed, whereas for the pure gBFKL pomeron
exchange we expect $n \sim $1.2 for the HERA kinematics
(\cite{GNZ95,GNZlong} and see below), for the pure pion exchange
$n\sim -1$. A comparison of the pion and pomeron exchanges
in \cite{HERApion} has shown that the
pion contribution becomes comparable to the pomeron contribution
at $x_{\Pom}\sim 0.1$. Consequently, the exponent $n$
must decrease from $n \sim$ 1.2 at small $x_{\Pom}$ down to
$n \sim -1$ in the pion exchange dominated region of larger
$x_{\Pom}$.

These two extreme cases demonstrate a potential sensitivity of
$n(x_{\Pom},\beta)$ to the non-pomeron exchanges. In a more accurate
treatment one must allow for the $f,\omega,\rho,A_{2}$
reggeon exchanges. In the color dipole gBFKL approach diffractive DIS is
controlled by predominantly soft interactions of large size color
dipoles  in the photon \cite{NZ92,NZ94,GNZ95}. For instance, it
has been argued that the so-called triple-pomeron coupling changes
little from real photoproduction, $Q^{2}=0$, to DIS at large
$Q^{2}$ \cite{GNZA3Pom}.
Similar dominance by soft pomeron interactions holds in
other popular models of diffractive DIS \cite{DLPom}.
Therefore, we can gain certain insight from the familiar
triple-Regge phenomenology of diffraction dissociation
of hadrons \cite{Kazarinov,FieldFox} and Regge fits to total cross
sections \cite{DLsigma}. The purpose of the present
communication is the quantitative evaluation
of the reggeon exchange contribution to $F^{D}$.
In conjunction with the kinematical $x_{\Pom}-\beta$
correlation, we find quite a strong impact of reggeon exchanges
on the effective exponent $n$  which is compatible with the
H1 finding. From the practical
point of view, better understanding of the
pomeron-reggeon-pion exchange content of diffractive DIS
is imperative for the interpretation of the large-$x_{\Pom}$
data to come soon from the Leading Proton Spectrometer of ZEUS.
We discuss simple tests of the reggeon-exchange mechanism of the
H1 effect and strategies for separation of the pure pomeron exchange.
We also point out the possibility of testing the predictions from
color dipole model of substantial intrusion of hard gBFKL exchange
to soft processes.
\section{Evaluation of the reggeon exchange parameters}

First, we recall the basics of the triple-Regge phenomenology
of hadronic diffraction.
The triple-Regge diagrams for $pp \rightarrow pX$ are shown
in Fig.~1 and give (for the sake of brevity we focus on $t=0$,
for detailed formulas see \cite{Regge,Kazarinov,FieldFox})
\arr
M^{2}
{ d\sigma^{D}
\over dt\,d M^{2} } =
{1 \over 16\pi}\left[
g_{\Pom}^{2}(t)|\xi_{\Pom}(t)|^{2}
\sigma_{tot}^{p\Pom}x_{\Pom}^{2(1-\alpha_{\Pom}(t))}+
g_{f}^{2}(t)|\xi_{f}(t)|^{2}\sigma_{tot}^{pf}
x_{\Pom}^{2(1-\alpha_{R}(t))}
\right.
\nonumber\\
+\left.2g_{\Pom}(t)g_{f}(t){\rm Re}[\xi_{\Pom}(t)\xi^{*}_{f}(t)]
\Sigma^{\Pom f}
x_{\Pom}^{2-\alpha_{\Pom}(t)-\alpha_{f}(t)}+...\right]
=
\nonumber \\
\phi_{\Pom}\sigma_{tot}^{p\Pom}+
\phi_{f}x_{\Pom}\sigma_{tot}^{pf}+\Phi_{\Pom f}x_{\Pom}^{1\over 2}
\Sigma^{\Pom f}...
\, ,
\label{eq:3}
\endarr
where $\phi_{\Pom}/x_{\Pom}$ and $\phi_{f}$ have the meaning of
fluxes of pomerons  and reggeons in the proton
\cite{Ingelman,DLPom} and $\sigma_{tot}^{p\Pom}$
and $\sigma_{tot}^{pf}$
have the meaning of the proton-pomeron and proton-reggeon total
cross sections, $\xi_{\Pom,f} = i-\cot({1\over 2}\pi\alpha_{\Pom,f})$
is the signature factor, the residues $g_{\Pom}$ and $g_{f}$ of
the pomeron and $f$ exchanges can be determined from
the crossing-even part of the $pp,\bar{p}p$ total cross
section
\beq
{1\over 2}(\sigma_{tot}^{pp}+\sigma_{tot}^{\bar{p}p})=
g_{\Pom}^{2}(0) +g_{f}^{2}(0)\left({s_{0}\over s}\right)^{1-\alpha_R}\, ,
\label{eq:4}
\endeq
where the standard but still arbitrary choice is $s_{0}=1$\,GeV$^{2}$.
In the $\Pom$-f interference term one encounters the amplitude of
forward diffraction dissociation of the pomeron into $f$ reggeon
and $\Sigma^{\Pom f} ={\rm Im}A(p\Pom \rightarrow pf)/M^{2}$.
For the sake of simplicity, formulas  (\ref{eq:3})
and (\ref{eq:4}) were written for $t=0$ and in the approximation
$\alpha_{\Pom}(0)=1$ and $\alpha_{R}(0)={1\over 2}$, the
former has been the common assumption (and deficiency) of all works
\cite{Kazarinov,FieldFox} on the triple-Regge analysis
of hadronic diffraction, we comment more on that below.
\vspace{1.0cm}
\begin{figure}[h]                                     
\begin{center}                                     
\epsfig{file=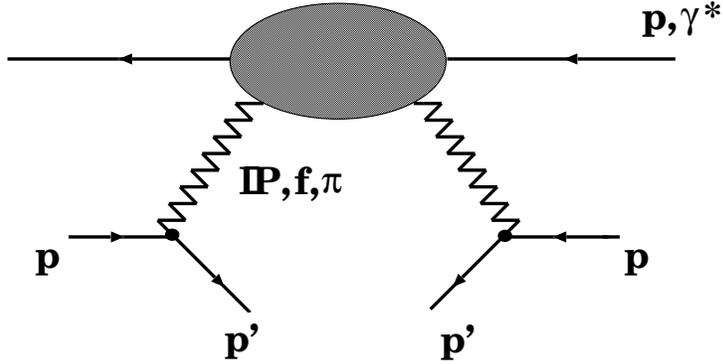,height=5.0cm}
\vspace{-.5cm}
\end{center}
\caption{\it Reggeon-exchange diagrams for $pp\rightarrow pX$.}
\label{phase-mcl-1}
\end{figure}

One must also include the numerically important pion exchange term
\beq
M^{2}
{ d\sigma^{D} (pp\rightarrow pX)
\over dt\,d M^{2} }
= {g_{\pi NN}^{2}x_{\Pom}^{2} \over (4\pi)^{2}}
{G_{\pi}^{2}(x_{\Pom},t)|t|\over (|t|+m_{\pi}^{2})^{2}}
\sigma_{tot}^{\pi N}\, ,
\label{eq:6}
\endeq
where $G_{\pi}(x_{\Pom},t)$ is the $\pi NN$ form factor.
The exchanged pions are not far off-mass shell and using the 
$\pi N$ total cross section for real pions leads to a
very good quantitative description of the related charge
exchange $pp \rightarrow nX$ (\cite{Sergeev,Zoller,Harald}
 and references therein). For diffractive DIS the similar
 substitution
\beq
\sigma_{tot}^{\pi N}\Longrightarrow \sigma_{tot}^{\gamma^{*}\pi}
={4\pi^{2} \alpha_{em}
\over Q^{2}}
 F_{2}^{\pi}(\beta,Q^{2})\,
\label{eq:7}
\endeq
with the Drell-Yan determinations of the pion structure function
is a viable approximation. For instance, it provides a 
parameter-free description of the experimentally observed
$\bar{u}-\bar{d}$ asymmetry in the proton (\cite{Harald,HaraldNA51}
and references therein).

The $f$-exchange contribution  to diffraction of protons is not
so well determined, the parameter $G_{ff\Pom} =
{1\over 8\pi}g_{f}^{2}(0)\sigma_{tot}^{pR}$
varies from 7.2\,mb(GeV)$^{-2}$ and 13.2\,mb(GeV)$^{-2}$ in
the two solutions of Kazarinov et al. \cite{Kazarinov} to
$\approx 30$\,mb(GeV)$^{-2}$ by Field and Fox \cite{FieldFox}.
The Regge fits to total cross sections are in much better
shape \cite{DLsigma} and give
$g_{f}^{2}(0)
\approx 80 mb$, for the $\omega$ exchange $g_{\omega}^{2}$
is one order in magnitude smaller, the residues for the $\rho,A_{2}$
are still smaller. Then, if we stick to the reggeon flux
normalization (\ref{eq:3}), the above cited determinations of
$G_{ff\Pom}$ would correspond to
\beq
\sigma_{tot}^{pf}={8\pi G_{ff\Pom}\over g_{f}^{2}(0)}=
\lambda_{f}\sigma_{tot}^{\pi N}=(1-4)mb \, ,
\label{eq:8}
\endeq
which is an order in magnitude smaller than the natural scale
$\sigma_{tot}^{\pi N}\approx 25$\, mb. Consequently, there
emerges a small parameter $\lambda_{f} =(0.04-0.15)$ and the
educated guess for the reggeon structure function in
the substitution (\ref{eq:7}) would be
\beq
F_{2}^{f}(\beta,Q^{2})\sim \lambda_{f}
F_{2}^{\pi}(\beta,Q^{2})\, .
\label{eq:9}
\endeq
Because the $\beta$ dependence of the reggeon structure
function is basically unknown, (\ref{eq:9}) must be
regarded as a useful benchmark evaluation.

 From the comparison of different triple-Regge fits
\cite{Kazarinov,FieldFox} one concludes that allowance for
the $\Pom$-f interference lowers the fitted value of $G_{ff\Pom}$.
In hadronic interactions the diffraction dissociation amplitudes
are strongly suppressed compared to elastic scattering amplitudes,
which suggests $\Sigma^{\Pom f}\ll \sigma_{tot}^{p\Pom},
\sigma_{tot}^{pf}$ and triple-Regge fits with weak $\Pom$-f
interference are more preferable. In diffractive DIS, the
$\Pom$-f interference term gives rise to an unusual off-diagonal
structure function associated with the imaginary part of
the $\gamma^{*}\Pom \rightarrow\gamma^{*}f$ 
forward scattering amplitude. It has been argued
that for hadronic targets such an off-diagonal structure function
is strongly suppressed because the quark number and momentum
integrals must vanish for the orthogonality of states \cite{HaraldNA51}.
Then, our educated guess is that in diffractive DIS the $\Pom$-f
interference must be negligibly small and the large $\lambda_{f}$
fit \cite{FieldFox} is preferred. 

There is one more reason for considering the larger $\lambda_{f}$.
Namely, an obvious deficiency
of the available triple-Regge fits \cite{Kazarinov,FieldFox} is
their assumption $\alpha_{\Pom}(0)=1$, whereas according to the
Donnachie-Landshoff fits to hadronic total cross sections
$\alpha_{\Pom}=1+\epsilon \approx 1.08$ is the more appropriate
one \cite{DLsigma}. For the Donnachie-Landshoff value of
$\alpha_{\Pom}(0)$ the pomeron contribution in the triple-Regge
expansion (\ref{eq:3}) will decrease $\propto x_{\Pom}^{-2\epsilon}$.
In order to reproduce the same observed diffraction cross section, 
this decrease of the pomeron contribution with increasing $x_{\Pom}$
ought to be compensated for by the enhancement of
the $f$-exchange contribution by about factor 2 compared to the
determinations in \cite{Kazarinov,FieldFox}. Therefore,
our educated guess for the $f$-reggeon contribution in the
triple-pomeron expansion (\ref{eq:10}) for diffractive DIS
is $\lambda_{f}\approx $0.3. Judging from evaluations of the
$Q^{2}$ dependence of the triple-pomeron coupling \cite{GNZA3Pom},
this estimate for $\lambda_{f}$ is good within the factor 2.

To conclude this discussion we mention that one must not
interpret the small $\lambda_{f}$ as a strong
suppression of the structure function of the strongly
off-mass shell reggeized $f$-meson because the
factor $\lambda_{f}$ can as well be reabsorbed into the
definition of the reggeon flux. Neither pomerons nor reggeons
can be treated as particles, the normalizations of
fluxes (\ref{eq:3}) and of the pomeron-particle and reggeon-particle
cross sections $\sigma_{tot}^{p\Pom},\sigma_{tot}^{pf}$ are
arbitrary, only the products $g_{\Pom}^{2}\sigma_{tot}^{p\Pom}$
and $g_{f}^{2}\sigma_{tot}^{pf}$ are well defined. For
instance, in the very definition of the Regge residue $g_{f}$ in
(\ref{eq:4}) there is a fundamental uncertainty with the
choice of $s_{0}$.

The pomeron exchange contribution to diffractive DIS has been
evaluated directly in terms of the color dipole gBFKL cross
section \cite{GNZ95} and the agreement with the HERA data on
$F^{D}$ is very good, see a detailed comparison between the
theory and experiment in \cite{ZEUSF2Pom}.

\section{The triple-Regge parameterization for diffractive DIS}

In (\ref{eq:3}) we focused on $t=0$. At HERA one rather
measures the $t$-integrated cross section, which
includes the charge exchange $\gamma^{*}p\rightarrow Xn$
on top of the $\pi^{0}$ exchange contribution $\gamma^{*}p
\rightarrow Xp$, there is also a certain admixture of
double diffraction, which is marginal
for the purposes of the present discussion, see the analysis
in \cite{DoubleD}. Within the present
uncertainty in the parameter $\lambda_{f}$ one can neglect
the difference in the $t$-dependence of the pomeron and
reggeon cross section. Then, our triple-Regge parameterization for
the observed mass spectrum is
\arr
(M^{2}+Q^{2})
{ d\sigma^{D} (\gamma^{*}\rightarrow X)
\over d M^{2} } =
{4\pi^{2} \alpha_{em}
\over Q^{2}}\Phi_{D}^{(3)}(x_{\Pom},\beta,Q^{2})=
~~~~~~~~~~~\nonumber\\
{4\pi^{2} \alpha_{em}
\over Q^{2}}\cdot
\left\{
{ \sigma_{tot}^{pp} G_{p}^{2}(m_{p}^{2}x_{\Pom}^{2})\over 16\pi B_{3\Pom}}
\left[(1+R_{LT})\phi_{\Pom}^{sea}(x_{\Pom})F_{sea}^{\Pom}(\beta,Q^{2})
+
{B_{3\Pom} \over B_{el}}\phi_{\Pom}^{val}
(x_{\Pom})F_{val}^{\Pom}(\beta,Q^{2})
\right.\right.\nonumber\\
\left. +\phi_{\Pom}^{L}
(x_{\Pom},Q^{2})F_{L,val}^{\Pom}(\beta,Q^{2})
\right]~~~~~~~~~~~~~~~~~
\nonumber\\+
\left.
{g_{f}^{2} G_{p}^{2}(m_{p}^{2}x_{\Pom}^{2})\over 8\pi B_{3\Pom}}
\lambda_{f}x_{\Pom}F_{2}^{\pi}(\beta,Q^{2})
+
3x_{\Pom}^{2}f_{\pi}(x_{\Pom})F_{2}^{\pi}(\beta,Q^{2})\right\}
\, ,~~~~~~~~~~~~~~~
\label{eq:10}
\endarr
where $G_{p}(q^{2})$ is the charge form factor of the proton
and $G_{p}^{2}(q{^2})$ gives an estimate for the survival of
the proton in the final state at the (longitudinal) momentum
transfer $q=m_{p}x_{\Pom}$.
For the numerical estimations, we take $B_{3\Pom}=6$\,GeV$^{-2}$
for the diffraction slope of the sea term and $B_{el}=2B_{3\Pom}$
for the valence term, $\sigma_{tot}^{pp}=40$\,mb,
$g_{f}^{2}/(8\pi B_{3\Pom}) \approx 1.3$. The pomeron contribution
has been described in \cite{GNZ95,GNZlong}. The parameterizations of
the valence and sea structure functions for $Q^{2}\sim 10$\,GeV$^{2}$
are $F_{sea}^{\Pom}(\beta,Q^{2})= 0.063(1-\beta)^{2}$,
$F_{val}^{\Pom}(\beta,Q^{2})=0.27\beta(1-\beta)$, the flux factors
\beq
\phi_{\Pom}(x_{\Pom}) =
\left({x_{0} \over x_{\Pom}}\right)^{p_{1}}\
\left({x_{\Pom} +p_{3} \over x_{0}+p_{3}}\right)^{p_{2}}
\label{eq:11}
\endeq
are normalized to unity at $x_{\Pom}=x_{0}=0.03$, for the
valence component of the pomeron $p_{1}=0.569, p_{2}=0.4895,
p_{3}=1.53\cdot10^{-3}$ ($\phi_{\Pom}^{val}(x_{\Pom})$ of the present
paper is $\phi_{\Pom}(x_{\Pom})$ of Ref. \cite{GNZ95}), and
for the sea component of the pomeron $p_{1}=0.741, p_{2}=0.586,
p_{3}=0.8\cdot 10^{-3}$ ($\phi_{\Pom}^{sea}(x_{\Pom})$ of the present
paper is $f_{\Pom}(x_{\Pom})$ of Ref. \cite{GNZ95}).
We included also the longitudinal cross section as calculated
and parameterized in \cite{GNZlong} (the parameterization
of \cite{GNZlong} was intended to reproduce the gross features
of $\Phi_{\Pom}^{L}$ and $F_{L,val}^{\Pom}$ only
at $x_{\Pom}\gsim 10^{-4}$); in the sea region $R_{LT}=
\sigma_{L}^{D}/\sigma_{T}^{D}=0.2$, the longitudinal
valence component
$F_{L,val}^{D}$ is small apart from the narrow region of
$\beta \gsim 0.8$, it takes over completely at $\beta \gsim 0.9$.

The pion contribution enters the diffractive
DIS as measured at HERA with the extra isospin factor 3,
because the both diffractive $\gamma^{*}p\rightarrow Xp$ and
charge-exchange $\gamma^{*}p
\rightarrow Xn$
channels are included in the observed cross section at the
present stage of the H1 and ZEUS detectors.
The pion structure function is borrowed from \cite{GRV},
a convenient parameterization for the flux of pions
good to  $\approx 10\%$ up to $x_{\Pom}\lsim 0.8$ (at
larger $x_{\Pom}$ this flux nearly vanishes anyway) is
\beq
f_{\pi}(x_{\Pom})x_{\Pom}^{2}={g_{\pi NN}^{2} x_{\Pom}^{2}
\over (4\pi)^{2}}\int_{m_{p}^{2}x_{\Pom}^{2}}^{\infty} d|t|
{G^{2}(t)|t|\over (|t|+m_{\pi}^{2})^{2}} \approx x_{\Pom}^{2}
0.66(1+5.7\sqrt{x_{\Pom}})(1-x_{\Pom})^{3.3}\, .
\label{eq:12}
\endeq

At small $\beta$ the $Q^{2}$ evolution of the pomeron and pion
structure functions must be similar \cite{NZ94,GNZ95} and
the gross features of $x_{\Pom}$ dependence in (\ref{eq:11})
must not change much with $Q^{2}$. The effects of non-GLDAP
evolution at $\beta\rightarrow 1$ \cite{GNZcharm,NZRoma} are
marginal for the purposes of the present exploratory study and
we present all the results for $Q^{2}=10$\,GeV$^{2}$.

\section{The reggeon-exchange driven factorization breaking:
the numerical estimates}

In Fig.~2 we show the $x_{\Pom}$ dependence of
$\Phi_{D}^{(3)}(x_{\Pom},\beta,Q^{2})$ defined by
Eq.~(\ref{eq:10}) for several values of
$\beta$. Unless specified otherwise, the reggeon contribution
is always evaluated for $\lambda_{f}=0.3$.
The pomeron-pion-reggeon content of $\Phi_{D}^{(3)}(x_{\Pom},\beta,Q^{2})$
varies with $\beta$ little. Because our parameterization for
$F_{sea}^{\Pom}(\beta,Q^{2})$ is flat at small $\beta$ whereas the
GRV pion structure function rises towards small $\beta$,
the pion and reggeon effects are enhanced at small $\beta$ slightly,
which may or may not survive for different pion and pomeron
structure function which for small $\beta$ are still a theoretical
guess. 
\vspace{1cm}
\begin{figure}[h]                                     
\begin{center}                                     
\epsfig{file=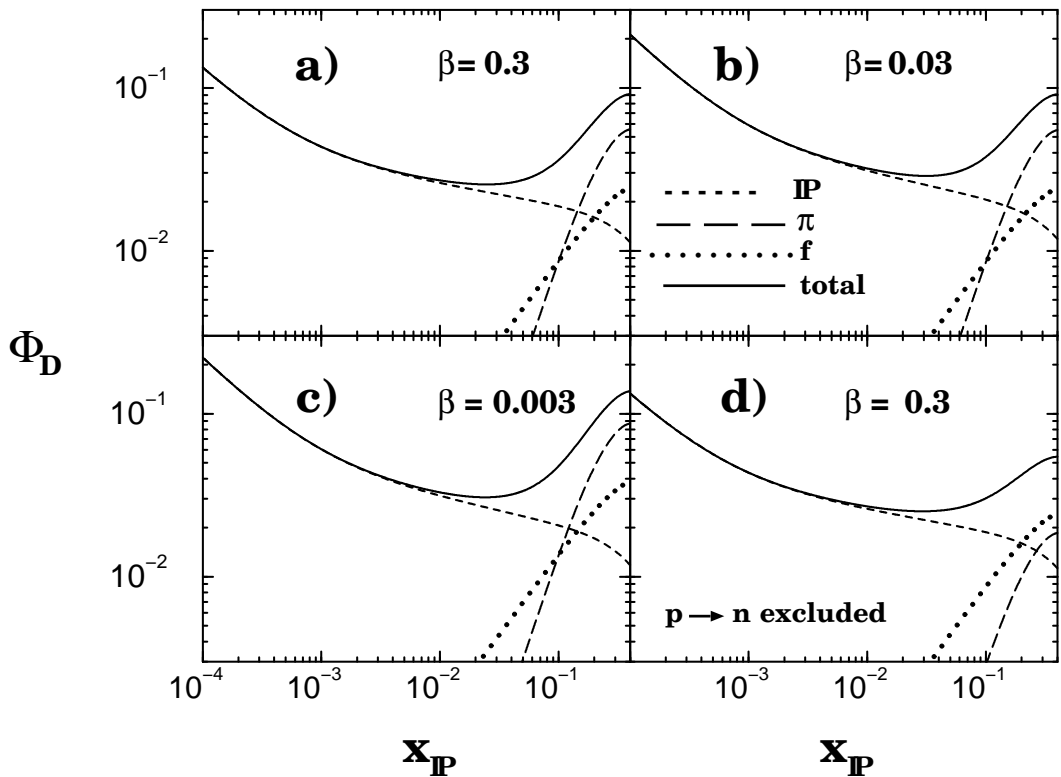,width=14.0cm} 
\end{center}
\vspace{-1cm}
\caption{\it  - The pomeron-reggeon-pion exchange decomposition
of the triple-Regge expansion for $\Phi_{D}^{(3)}(x_{\Pom},\beta,Q^{2})$.
The $f$-exchange is evaluated for $\lambda_{f}=0.3$. The boxes
(a), (b), (c) are for $\beta=0.3, 0.03, 0.003$, respectively.
The box (d) is for the LPS trigger which excludes the contribution
from the charge exchange reaction $\gamma^{*}p \rightarrow Xn$.}

\end{figure}
Typically,
the reggeon contribution becomes noticeable at $x_{\Pom}\sim 0.01$,
the pion contribution dominates at $x_{\Pom} \gsim 0.1$. The
combined reggeon and pion contributions substantially alter the
trend of the $x_{\Pom}$ dependence already at
$x_{\Pom} \gsim 0.02$. Their effect is best seen in the
exponent of the local $x_{\Pom}$ dependence
$$
n(x_{\Pom},\beta)= 1-{d \log \Phi_{D}^{(3)}(x_{\Pom},\beta,Q^{2})
\over d\log x_{\Pom}}\,
$$
shown in Figs.~3 and 4. In Fig.~3 we show this exponent
$n_{\Pom}(x_{\Pom},\beta)$ for the
pure pomeron exchange and for the unconstrained $\beta$ and
$x_{\Pom}$ dependence. In the specific color dipole model
\cite{NZZ94,NZHera} the rightmost $j$-plane singularity of
the gBFKL pomeron has an intercept $\alpha_{\Pom}(0)=1.4$,
and at very high energies and/or very small $x,x_{\Pom}$ all
cross sections, for soft and hard processes alike, must exhibit
the universal $\propto s^{\alpha_{\Pom}(0)},x^{-\alpha_{\Pom}(0)}$
behavior. Indeed, at larger $x_{\Pom}$ the nonperturbative
soft pomeron dominates and $n$ is small, with the rising
contribution from the gBFKL pomeron  at very small $x_{\Pom}$
the exponent
$n(x_{\Pom},\beta)$ tends to $n=2\alpha_{\Pom}(0)-1
=1.8$, although very small $x_{\Pom}$ beyond the
HERA range is needed to reach this limiting value
\footnote{The exponents $p_{1}$ for the sea and valence
fluxes are close to but still
unequal to $2(\alpha_{\Pom}(0)-1)=0.8$, because the simple
parameterization (\ref{eq:11}) was intended to describe
the flux functions only at $x\gsim 10^{-5}$ with an emphasis
on the still larger values of $x_{\Pom}$ accessible at
HERA.} \cite{NZHera}. For the soft pomeron dominated mechanisms
of diffractive DIS $n(x_{\Pom},\beta)$ is flat vs. $x_{\Pom}$
\cite{DLPom}.
Notice, that variations of $n(x_{\Pom},\beta)$ with $x_{\Pom}$
are quite substantial, stronger than variations with $\beta$.
The  latter derives from the factorization
breaking difference  of $\phi_{\Pom}^{val}(x_{\Pom})$ and
$\phi_{\Pom}^{sea}(x_{\Pom})$ and from the dominance of the
longitudinal cross section at $\beta \gsim 0.9$,
which is the higher twist effect and is less important at
larger $Q^{2}$ \cite{GNZlong}.
\vspace{1cm}
\begin{figure}[h]                                     
\begin{center}                                     
\epsfig{file=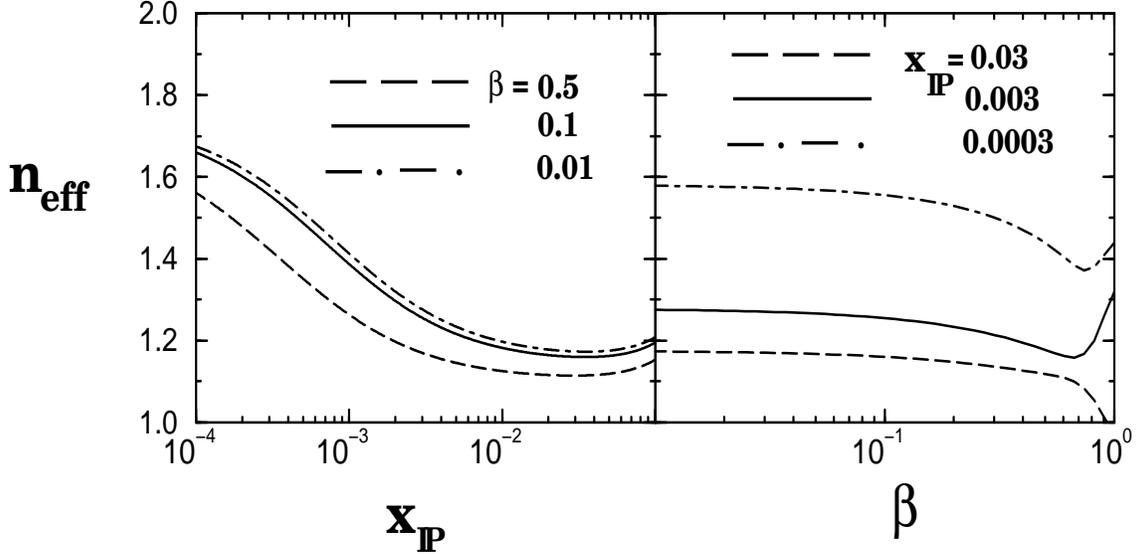,width=15.0cm,height=7.5cm}
\vspace{-1.0cm} 
\end{center}
\caption{\it The exponent $n_{\Pom}(x_{\Pom},\beta)$ for the pure pomeron exchange.
In the left box we show only $\beta > 0.01$, at smaller $\beta$
the exponent $n_{\Pom}(x_{\Pom},\beta)$ levels off.}
\end{figure}
In Fig.~4 we present predictions from the full triple-Regge
expansion (\ref{eq:10}). In the typical experimental situation
the smaller values of $\beta$ imply the larger values of
$x_{\Pom}$ which enter the fit $\Phi^{D}\propto x_{\Pom}^{1-n}$.
Then,
the gross features of this $x_{\Pom}$-$\beta$ correlation are
reproduced by
\beq
\langle x_{\Pom} \rangle \sim {\langle x(\beta) \rangle \over \beta}\, .
\label{eq:14}
\endeq
\vspace{0.5cm}
\begin{figure}[h]                                     
\begin{center}                                     
\epsfig{file=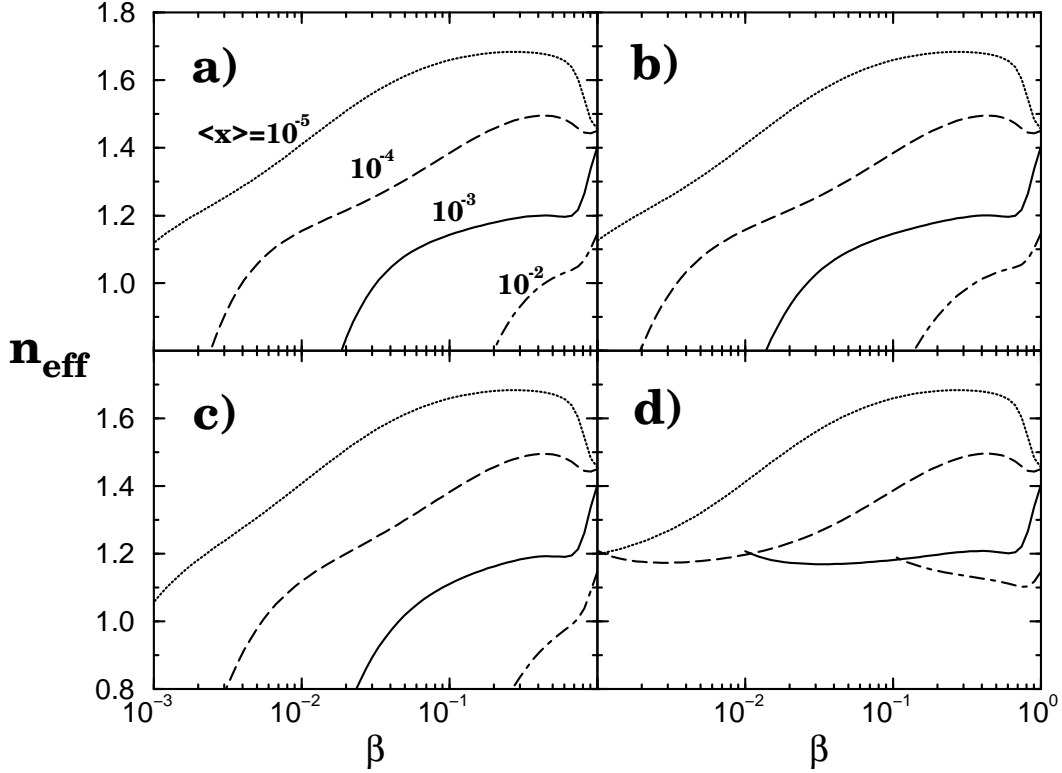,width=14.0cm} 
\vspace{-1.cm}
\end{center}
\caption{\it- The exponent $n_{eff}(\langle x \rangle, \beta)$
evaluated subject to the correlation
(\ref{eq:14}) for different values of $\langle x \rangle$ which
are the same for all the boxes:
(a) the results for $\lambda_{f}=0.3$ and no forward nucleon
trigger; (b) the same as (a) but for the LPS trigger which
excludes the $\gamma^{*}p \rightarrow Xn$ contribution; (c)
the same as (a) but for $\lambda_{f}=0.6$; (e) the exponent
$n_{eff}(\beta)$ for the pure pomeron exchange.}
\end{figure}
The somewhat different range of $x$ is spanned at different
$\beta$, for the purposes of our crude estimates we can
take $\langle x (\beta)\rangle= 10^{-3}$ \cite{H1F2Pom,ZEUSF2Pom,H1Roma}.
The H1 determinations of the exponent $n$ correspond to
$n_{eff}(\langle x\rangle,\beta)=
n(x_{\Pom}={\langle x \rangle \over \beta},\beta)$
and the $x_{\Pom}$ dependence of $n(x_{\Pom},\beta)$
shown in Fig.~3 transforms into the effective $\beta$ dependence
of $n_{eff}(\langle x \rangle,\beta)$ shown in Fig.~4. 
In order to see the impact of the $x_{\Pom}$-$\beta$ correlation,
focus for simplicity on $\beta \ll 1$ and suppress the pion effects.
For the GRV pion structure function
$F^{\pi}(\beta,Q^{2}) \approx CF_{se}^{\Pom}(\beta,Q^{2})$ 
with $C \sim 10$ and the both structure functions are 
approximately flat for $\beta \gsim 10^{-2}-10^{-3}$. The
Donnachie-Landshoff fits five
$g_{\omega}^{2} /\sigma_{tot}^{pp} \approx 2$. Then
$\Phi_{D}^{(3)} \propto \phi_{\Pom}^{sea}(x_{\Pom}) + 4C\lambda_{f}
x_{\Pom}^{1-\alpha_{R}}$ and
\bea
n_{eff}(\langle x  \rangle ,\beta )=
n_{\Pom}(x_{\Pom},\beta)
{\phi_{\Pom}^{sea}(x_{\Pom}) \over 
\phi_{\Pom}^{sea}(x_{\Pom}) + 4C\lambda_{f}x_{\Pom}^{1-\alpha_{R}}}
\nonumber\\
-\left({ \langle x \rangle \over \beta}\right)^{1-\alpha_{R}}
{ 4C\lambda_{f}(1-\alpha_{R}) \over 
\phi_{\Pom}^{sea}(x_{\Pom}) + 4C\lambda_{f}
x_{\Pom}^{1-\alpha_{R}}} \, ,
\label{eq:exponent}
\eea
where $x_{\Pom}=\langle x \rangle /\beta$ is understood everywhere,
we used that explicitly in front of the major correction term.
Evidently, the $f$-reggeon contribution depletes $n_{eff}$ and the
smaller is $\beta$ the stronger is the depletion of $n_{eff}$.
The effect of pions is similar.
In Fig.~4a we show $n_{eff}(\langle x \rangle,\beta)$ for the
triple-Regge expansion (\ref{eq:10})
with $\lambda_{f}=0.3$. In Fig.~4b we also show $n(x_{\Pom},\beta)$
for diffractive DIS measured with the Leading Proton Spectrometer
(LPS) trigger which excludes the charge exchange
$\gamma^{*}p \rightarrow Xn$ signal, i.e.,
removing the factor 3 from the pion contribution in (\ref{eq:10}).
Fig.~4c is for the enhanced reggeon exchange, $\lambda_{f}=0.6$.
Finally, Fig.~4d shows the exponent $n_{eff}(\beta)$ for
the pure pomeron exchange
evaluated subject to the same $x_{\Pom}$-$\beta$ correlation
(\ref{eq:14}). The impact of the $x_{\Pom}$-$\beta$ correlation
is non-negligible for the pure pomeron exchange too. First, the
rise of $n(x_{\Pom},\beta)$ with the decreasing $\beta$ which
is clearly seen in Fig.~3 , transforms
into flattening of $n_{eff}(\langle x \rangle =0.001,\beta)$
and for smaller values of
$\langle x \rangle$ into a substantial depletion
of $n_{eff}(\langle x \rangle,\beta)$ at small $\beta$ starting from
a fairly large values of $\beta$.

\begin{figure}[h]                                     
\begin{center}                                     
\epsfig{file=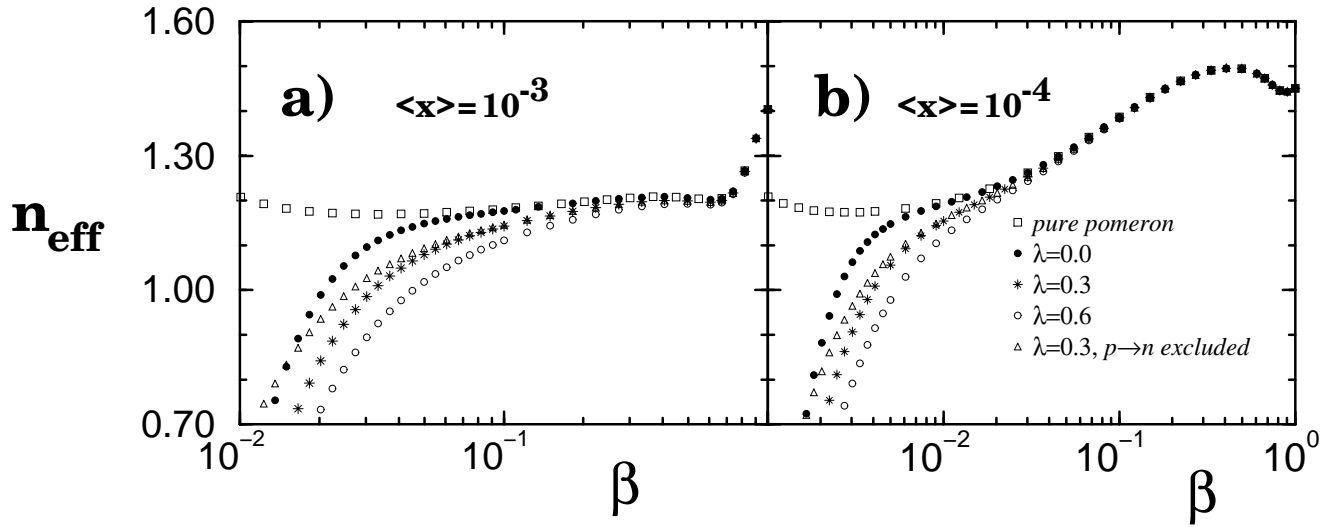,height=7.cm} 
\vspace{-1.cm}
\end{center}
\caption{\it- (a) A comparison of $n_{eff}(\langle x \rangle,\beta)$ evaluated at
$\langle x \rangle =10^{-3}$ for the pure
pomeron exchange (squares) and for different assumption on the
reggeon and pion exchange; (b) the same as (a) but for
$\langle x \rangle =10^{-4}$.
Notice, that the $\beta$ scales for the two boxes
are different.}
\end{figure}

A more detailed comparison of our results for the pure pomeron
exchange and different $f$ and $\pi$-exchange background is
presented in Fig.~5. In Fig.~5a we take $\langle x \rangle = 10^{-3}$
which is typical of the data taking in the H1 experiment. The
allowance for the $f$-reggeon and pion exchanges leads to a
substantial depletion of $n_{eff}(\langle x \rangle,\beta)$ at
small $\beta$ compared to the pure pomeron exchange; the
form of the depletion is similar to that reported by H1
\cite{H1Roma}. For
$\langle x \rangle=10^{-3}$ we find that the pion contribution
affects the exponent $n$ only at $\beta \lsim 0.05$, the
depletion at larger $\beta$ predominantly comes from the $f$-exchange.
A careful treatment of the $x_{\Pom}$-$\beta$ correlation
which depends on the experimental acceptances is needed to
draw quantitative conclusions on the value of $\lambda_{f}$.
One point is clear, though: already in the presently
available data the points at largest $\beta$ are free of the pion and
reggeon effects and measure the exponent $n_{eff}(\langle x\rangle,
\beta)$ for the pure pomeron exchange. Notice the spike at 
$\beta \rightarrow 1$ in Fig.~5a, which comes from the
dominance of the longitudinal structure function at
$\beta \gsim 0.9$ \cite{GNZlong}.

\section{How to separate the pure pomeron exchange?}

The chief purpose of experiments on diffractive DIS is a
study of the pomeron exchange and one would like to
exclude the reggeon and pion contributions. To this end,
recall that even the pion structure function is basically
unknown at $x\lsim 0.2$ and the available parameterization
for $F_{2}^{\pi}$ differ markedly \cite{GRV,Sutton}.
Nevertheless, the pion contribution to $F^{D}$ can be separated
without much problems. Whereas the both diffractive
$p\rightarrow p$ and charge-exchange $p\rightarrow n$
channels do contribute to the present data,
the LPS trigger will select the diffractive
$p\rightarrow p$ channel and
lower the pion contribution to the triple-Regge expansion
(\ref{eq:10}) by the factor 3. The effect of
the LPS trigger on  $n(\langle x \rangle,\beta)$
is shown in Figs.~4b and 5. Furthermore, the Forward
Neutron Calorimeters now in operation at the both ZEUS
\cite{ZEUSneutron} and H1 \cite{H1neutron} allow a direct
measurement of the charge exchange reaction
$\gamma^{*}p \rightarrow Xn$ and then the isospin
relation $d\sigma(\gamma^{*}p \rightarrow Xp)=
{1\over 2}d\sigma(\gamma^{*}p \rightarrow Xn)$ can be used.
The LPS and FNC triggers allow a direct determination of the
combined effect of all the isovector $\pi,\rho,A_{2}$ exchanges.

\begin{figure}[h]                                     
\begin{center}                                     
\epsfig{file=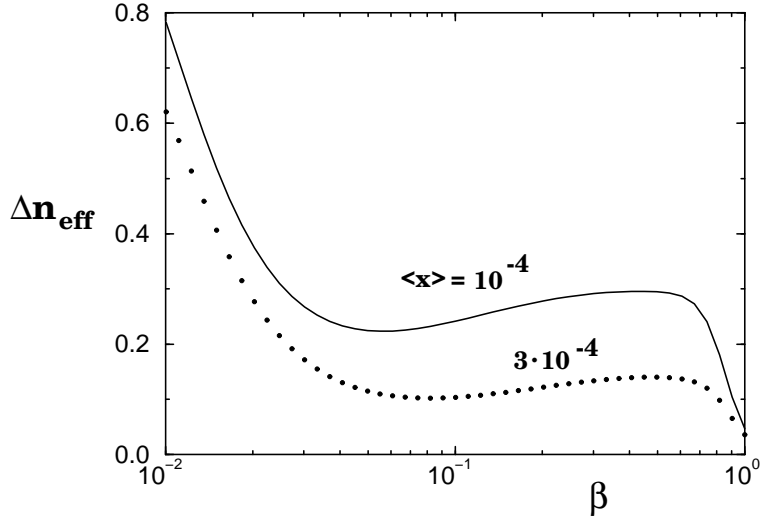,width=10.0cm} 
\vspace{-1.cm}
\end{center}
\caption{\it The change of $n_{eff}(\langle x \rangle,\beta)$ with
$\langle x \rangle$ shown in the form of $\Delta n_{eff}=
n_{eff}(\langle x \rangle,\beta)-
n_{eff}(\langle x \rangle=0.001,\beta)$.}
\end{figure}

An independent direct evaluation of the poorly known $f$ exchange
contribution is not possible. The $f$ contribution can be
minimized going to smaller values of $\langle x\rangle$
and/or $Q^{2}$, which for fixed $\beta$ implies smaller
values of $x_{\Pom}$. In Fig.~4 we show the effect of
changing $\langle x\rangle $, the smaller is $\langle x\rangle$
the weaker is the departure from predictions for the pure
pomeron exchange shown in Fig.~4d. This point 
is clear from equation (\ref{eq:exponent}) and is further
demonstrated in Fig.~5b, in which we compare $n_{eff}(
\langle x \rangle=10^{-4},\beta)$
evaluated for the pure pomeron exchange and with
allowance for the reggeon and pion contributions.
Going to smaller $\langle x \rangle$ entails smaller $Q^{2}$,
of course. Here we wish to emphasize that at least in the
color dipole gBFKL dynamics properties of the exchanged
pomerons do not vary with $Q^{2}$ as soon as $Q^{2}
\gsim $(2-3)\,GeV$^{2}$ \cite{GNZA3Pom}. The principal effect
of lowering $\langle x \rangle $ and $x_{\Pom}$ will be the
enhancement of the gBFKL pomeron contribution to the flux factors.
The resulting increase of $n(x_{\Pom},\beta)$ at small $x_{\Pom}$
shown in Fig.~3 entails a substantial increase of
$n_{eff}(\langle x \rangle,\beta)$ when $\langle x \rangle$
is lowered from $\langle x \rangle =10^{-3}$ down to
$\langle x \rangle =10^{-4}$. This very specific
prediction from the color dipole gBFKL pomeron is best
illustrated in Fig.~6, where we show $\Delta n_{eff}=
n_{eff}(\langle x \rangle,\beta)-
n_{eff}(\langle x \rangle=0.001,\beta)$ for two values
of $\langle x \rangle$. The accuracy of experimental
determinations of $n_{eff}$ is already sufficiently high
for testing this gBFKL prediction for $\Delta n_{eff}$.

\section{Conclusions}

The triple-Regge phenomenology is called upon for the
quantitative interpretation of diffractive DIS as measured
in the HERA kinematical domain. Based on the triple-Regge
analysis of hadronic diffraction, we formulated expectations
for the flux and structure function of secondary reggeons
in diffractive DIS. The $f$-reggeon and pion exchange are
the two prominent contributions. We argued that the $\Pom$-f
interference contribution must be small. We find a
substantial non-pomeron background, which is strongly
enhanced at small $\beta$ because of the kinematical
$x_{\Pom}$-$\beta$ correlation. The subasymptotic gBFKL
effects also contribute strongly to the observed $\beta$
dependence of $n_{eff}(\langle x \rangle,\beta)$.
The emerging pattern of factorization
breaking is consistent with the preliminary data from the H1
experiment. The proposed interpretation of the H1 effect
can be tested eliminating the pion (and isovector reggeons in
general) background to the pomeron exchange either using the
LPS trigger or measuring $\gamma^{*}p \rightarrow Xn$ with the
forward neutron detectors. Our analysis shows that the large-$\beta$ 
results for $n_{eff}(\beta)$ from H1, ZEUS  are already 
free of the reggeon and pion effects and biases for the $x_{\Pom}$-$\beta$ 
correlation and probe the pure pomeron exchange. The $f$-reggeon
contribution will be significantly lowered and will be
marginal at $\beta \gsim 0.03$ in a data sample taken
at $\langle x \rangle \sim 10^{-4}$. A substantial rise
of $n_{eff}(\beta)$ by $\approx$0.25 over the whole range
of $\beta$ studied experimentally when $ \langle x \rangle$
is lowered from $\langle x \rangle \sim 10^{-3}$ down to
$\langle x \rangle\sim 10^{-4}$ offers a stringent test
of the color dipole gBFKL approach.

After this work was completed, J.Dainton informed us of
a related analysis of the H1 data in terms of the
$f$-reggeon exchange \cite{H1Warsaw}.
\\

{\bf Acknowledgments:} Thanks are due to J.Dainton, 
A.Mehta, and J.Phillips for
helpful discussions and communications on the H1 data.
NNN thanks Prof. U.Mei{\ss}ner for the hospitality
at the Inst. Theor. Kernphysik of the Univ. of Bonn. The work of NNN 
is supported by the DFG grant ME864/13-1.

\pagebreak

\noindent

\end{document}